\documentclass[fleqn,twoside]{article}
\usepackage{espcrc2}

\usepackage{graphicx}
\usepackage[figuresright]{rotating}

\newcommand{\AmS}{{\protect\the\textfont2
  A\kern-.1667em\lower.5ex\hbox{M}\kern-.125emS}}

\newcommand{\lsim}{\mathrel{\rlap{\lower4pt\hbox{\hskip0pt$\sim$}} 
\raise1pt\hbox{$<$}}}           
\newcommand{\gsim}{\mathrel{\rlap{\lower4pt\hbox{\hskip0pt$\sim$}} 
\raise1pt\hbox{$>$}}}           

\title{Confinement, DCSB, Bound States,
and the Quark-Gluon Vertex
}

\author{A.\ H\"oll,\address[ANL]{%
Physics Division, Argonne National Laboratory, Argonne IL 60439, 
USA} 
A.\ Krassnigg\,\addressmark[ANL]
\ and C.\,D.\ Roberts\,\addressmark[ANL]$^{,}$\address{%
Fachbereich Physik, Universit\"at Rostock, D-18051 Rostock, Germany}}

\begin{document}

\begin{abstract}
Aspects of the dressed-quark-gluon vertex and their role in the gap and Bethe-Salpeter equations are briefly surveyed using an intuitive model.  The model allows one to elucidate why a linear extrapolation to the chiral limit of extant lattice data on the dressed-quark mass-function overestimates this function and hence the value of the vacuum quark condensate.  The diagrammatic content of the vertex described is explicitly enumerable.  This property is essential to the symmetry preserving study of bound state properties.  It facilitates a realistic analysis of vector and pseudoscalar meson masses, and also allows the accuracy of standard truncations to be gauged.  The splitting between vector and pseudoscalar meson masses is observed to vanish as the current-quark mass increases.  That argues for the mass of the pseudoscalar partner of the $\Upsilon(1S)$ to be above $9.4\,$GeV.  Moreover, in this limit the rainbow-ladder truncation provides an increasingly accurate estimate of a bound state's mass. 
\vspace{1pc}
\end{abstract}

\maketitle

\setcounter{footnote}{0}
It is a longtime prediction of Dyson-Schwin\-ger equation (DSE) studies that the Schwinger functions which characterise the propagation of QCD's elementary excitations are strongly modified at infrared momentum scales
\cite{cdragw}.  This capacity of asymptotically free theories was elucidated in Refs.\ \cite{lanepolitzer}.  Simulations of lattice-QCD also provide access to QCD's Schwinger functions, and studies of the quenched theory yield gluon \cite{latticegluon2} and quark \cite{bowman2} propagators that are in semi-quantitative agreement with DSE calculations.  It was anticipated \cite{hawesmaris} and shown \cite{mandarlattice} that the lattice gluon propagator can generate a realistic quark propagator via QCD's gap equation only so long as the quark-gluon vertex exhibits material infrared enhancement.  Such behaviour can arise as a consequence of multiplicative renormalisability of the gap equation \cite{bloch} and may be evident in lattice estimates of this three-point function \cite{latticevertex3}.

These and other features of QCD's elementary Schwinger functions, and also properties of higher-$n$-point functions and thus bound states, can be illustrated by working with a simple model for the quark-gluon vertex; viz., 
\begin{eqnarray}
\nonumber \lefteqn{
\Gamma_\mu^{\cal C}(k_+,k_-) =  \gamma_\mu  -\, {\cal C}\,\frac{4}{3}\, \int \! \frac{d^4 \ell}{(2\pi)^4}\, {\cal D}_{\rho\sigma}(p-\ell)
}\\
&& \times \,
\gamma_\rho S(\ell_+) \Gamma_\mu^{\cal C}(\ell_+,\ell_-)S(\ell_-) 
\gamma_\sigma \,, \label{G2C} 
\end{eqnarray}
wherein $k_\pm=k \pm p/2$, $S(\ell)$ is the dressed-quark propagator and ${\cal D}_{\rho\sigma}(\ell)$ is a dressed-gluon interaction.\footnote{A full motivation, and a discussion of issues such as renormalisation, is available in Ref.\,\protect\cite{mandarvertex}.  It is particularly significant that the diagrammatic content of this model is explicitly enumerable \cite{axeltrunc,detmold}.  NB. A rainbow-ladder truncation follows from ${\cal C}=0$ in Eq.\,(\protect\ref{G2C}).}  The quantity ${\cal C}$ is a global parameter.  It ensures that Eq.\,(\ref{G2C}) can yield a solution whose properties mimic those of the true vertex in QCD while avoiding the need to introduce an \textit{Ansatz} for the three-gluon vertex.  The model described by Eq.\,(\ref{G2C}) possesses material illustrative capacity.  For example, an efficacious \textit{Ansatz} for the three-gluon vertex was recently introduced \protect\cite{mandarnew}.  The results obtained therein are semi-quantitatively in agreement with those obtained in Ref.\,\cite{mandarvertex}, which are based on Eq.\,(\protect\ref{G2C}).  Nevertheless, adapting this \textit{Ansatz} to the needs of Ref.\,\cite{mandarvertex} could obviate the need for the parameter ${\cal C}$.

The gap equation is 
\begin{eqnarray}
\nonumber
\lefteqn{ S(k)^{-1} = i\gamma\cdot k + m }\\
&& + \frac{4}{3}\int \! \frac{d^4 \ell}{(2\pi)^4}\, {\cal D}_{\mu\nu}(k-\ell) \gamma_\mu S(\ell) \Gamma^{\cal C}_\nu(\ell,k)\,.
\label{gap}
\end{eqnarray}
Equations (\ref{G2C}) and (\ref{gap}) form a closed system once ${\cal D}_{\rho\sigma}(\ell)$ is specified.  A particularly useful confining model is \cite{mn83}
\begin{equation} 
\label{mnmodel} {\cal D}_{\mu\nu}(\ell):=  
\left(\delta_{\mu\nu} - \frac{\ell_\mu \ell_\nu}{\ell^2}\right) (2\pi)^4\, {\cal G}^2 \, \delta^4(\ell)\,. 
\end{equation} 
The constant ${\cal G}$ sets the mass-scale.  Equation (\ref{mnmodel}) can be used to complete a definition of the dressed-ladder quark-antiquark scattering kernel 
\begin{eqnarray}
\nonumber\lefteqn{
K^{tu}_{rs}(q,k;P)}\\
&& = - \,{\cal D}_{\mu\nu}(k-q) \left[\gamma_\mu \frac{\lambda^a}{2}\right]_{ts} \, \left[\gamma_\nu \frac{\lambda^a}{2}\right]_{ru} \!. \label{ladderK}
\end{eqnarray}
In this case Eqs.\,(\ref{mnmodel}), (\ref{ladderK}) describe an interaction that may be rendered as an integrable regularisation of $1/k^4$ \cite{mm97}.  This interpretation, when combined with the result that, in the absence of anomalous infrared divergences, the renormalisation-group-improved (RGI) ladder truncation of $(S\otimes S)K$ is exact in a heavy-quark--heavy-antiquark Bethe-Salpeter equation, is consistent with a linear potential between static sources at long-range.  

The model thus constructed is very useful.  It facilitates an algebraic analysis of what would otherwise be coupled integral DSEs and hence the rapid development of insight.  More laborious studies with refined model kernels in the coupled integral equations may then follow and augment this understanding; e.g., as Ref.\,\cite{mandarnew} complements Ref.\,\cite{mandarvertex}.

It is straightforward to solve Eqs.\,(\ref{G2C}), (\ref{gap}) using Eq.\,(\ref{mnmodel}).  The properties of the solution are described at length in Ref.\,\cite{mandarvertex}.  To recapitulate in part, with Eq.\,(\ref{mnmodel}) one obtains information about 
\begin{eqnarray}
\nonumber \lefteqn{
\Gamma^{\cal C}_\mu(p,p) = \gamma_\mu \, \alpha_1^{\cal C}(p^2) }\\
&&+ \gamma\cdot p \,p_\mu\, \alpha_2^{\cal C}(p^2) - i p_\mu \, \alpha_3^{\cal C}(p^2)\,.
\end{eqnarray}
A value of ${\cal C}=0.51$ yields a qualitatively and semi-quantitatively accurate description of available lattice data on the dressed-quark-gluon vertex \textit{except} for $\alpha_2(p^2)$: in the infrared, the model result for $\alpha_2^{\cal C}(p^2)$ is approximately an order of magnitude smaller than a recent lattice estimate~\cite{latticevertex3}.  It is noteworthy that this pattern is repeated in Ref.\,\cite{mandarnew}, which reports results for $\alpha_{1,3}(p^2)$ that are in quantitative agreement with the lattice estimates but a result for $\alpha_2(p^2)$ that disagrees markedly with the available lattice data.  It is currently not possible to understand the lattice data on $\alpha_2(p^2)$ using continuum methods.

The solution of Eqs.\,(\ref{G2C})--(\ref{mnmodel}) also gives the dressed-quark propagator
\begin{equation}
S(k) = Z(k^2)/[i\gamma\cdot k + M(k^2)]\,,
\end{equation}
where $Z(k^2)$ is the quark wave-function renormalisaton and $M(k^2)$ is the quark mass-function.  Lattice data for these functions is available at nonzero current-quark masses on the spacelike $k^2$ domain whereat Eqs.\,(\ref{G2C})--(\ref{mnmodel}) again give results in qualitative and semi-quantitative agreement.  

It has been observed \cite{mandarlattice} that a linear extrapolation of lattice data to the chiral limit overestimates the mass-function calculated in the chiral limit using the gap equation, particularly on the domain $0.5 \lsim k^2 \,(\mbox{GeV}^2)\,\lsim 2.0$.  This fact may be understood by considering a chiral susceptibility; namely, define $M(k^2,m)$ to be the dressed-quark mass-function evaluated with a current-quark mass, $m$, and explore the behaviour of
\begin{equation}
\label{Chim} 
\chi_m(k^2) := \frac{\partial}{\partial m} \, M(k^2,m)\,.
\end{equation}

\begin{figure}[t] 
 
\centerline{\includegraphics[width=0.45\textwidth]{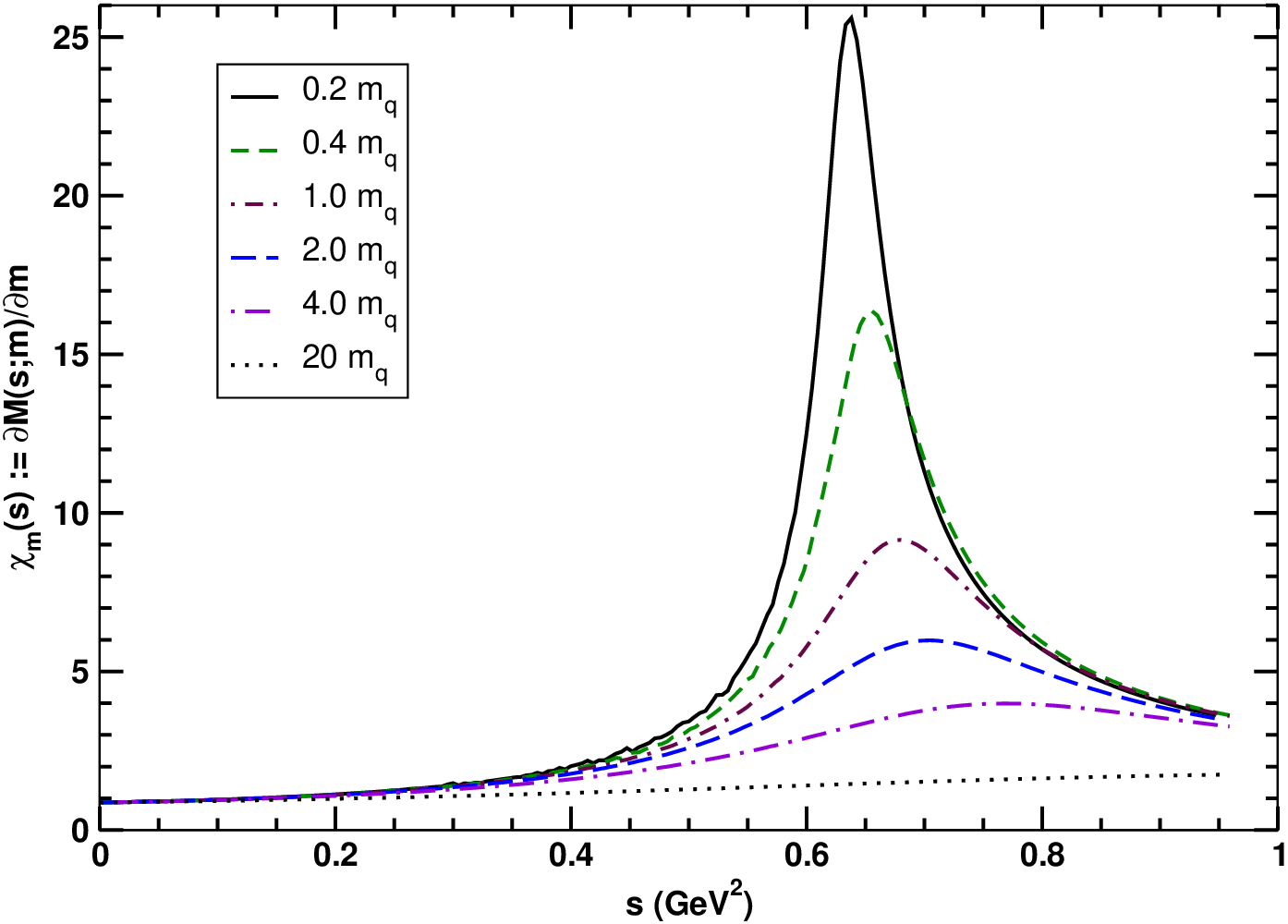}}
 
\caption{\label{Fig1} Chiral susceptibility, Eq.\,(\protect\ref{Chim}), plotted against $s=k^2$ for a range of values of the current-quark mass. ($m_q=7.5\,$MeV.  Results calculated using model explained in the text, which incorporates a dressed quark-gluon vertex.)}
\end{figure}

The susceptibility obtained using Eqs.\,(\ref{G2C})--(\ref{mnmodel}); viz., the model described fully in Ref.\,\cite{mandarvertex}, is plotted in Fig.\,\ref{Fig1}.  It is positive, which is intuitively sensible because in QCD an increase in the current-quark mass should also increase the size of the dressed-quark mass-function at every $k^2$.

The behaviour evident in Fig.\,\ref{Fig1} will be familiar to those acquainted with analyses; e.g., of the chiral symmetry restoring transition in quantum field theory at nonzero temperature, such as those in Refs.\,\cite{arnesusc}, as soon as one appreciates that the momentum argument, $s=k^2$, is behaving as a surrogate for the extrinsic variable temperature.  The $s$-dependence is then typical, with the peak increasing rapidly in height as $m\to 0$; i.e., as the external source for chiral symmetry breaking diminshes.  

To understand this response, remember that the chiral susceptibility is the derivative of the mass-function with respect to the explicit chiral symmetry breaking mass.  The typical mass scale associated with dynamical chiral symmetry breaking (DCSB) is $M_\chi$, with $M_\chi^3  \sim -\langle\bar q q\rangle$ ($M_\chi \sim 0.13\,$GeV in this model).  For $m \gg M_\chi$, explicit chiral symmetry breaking dominates for all $s$, so that $M(s,m) \approx m$, and hence $\chi_m(s) \approx\,$const.  A linear extrapolation of lattice data on $M(s,m)$ from one large current-quark mass to another should therefore be reliable.

For $m\sim M_\chi$, $\chi_m(s)$ varies with $s$ as this variable passes from the ultraviolet domain, on which the explicit current-quark mass dominates, into the infrared, where the DCSB mass-scale controls the behaviour of $M(s,m)$.  The transition between these domains is marked by the appearance of a peak in the susceptibility, whose height indicates the sensitivity of the mass-function to changes in the current-quark mass.

For $m< M_\chi$, the scale of the dressed-quark mass-function is still determined by $M_\chi$ at small $s$ and hence $\chi_m(s)$ remains approximately constant.  Indeed, the domain of small $s$ on which $\chi_m(s)$ is constant truly defines the domain of infrared momenta: $s\lsim 0.5\,$GeV$^2$ in this case.  With increasing $s$ one enters the perturbative domain; namely, the region on which a nonzero mass-function\footnote{In QCD this translates into a mass-function that does not exhibit an essential singularity in the running coupling.} is only possible for $m\neq 0$.  Hence, for small $m$, the mass-function must evolve rapidly with $s$, and its size at a given value of $s$ will depend sensitively on $m$.  Consequently, $\chi_m(s)$ is large.  Further into the ultraviolet, one again has $M(s,m)\approx m$ and hence $\chi_m(s) \approx\,$const.\ once more.  

The region on which $\chi_m(s)$ is large demarcates the domain of transition between infrared and ultraviolet momenta: $0.5\lsim s \,(\mbox{GeV})^2 \lsim 1.0$.  It is thus for $s\gsim 0.5\,$GeV$^2$ that the size of the vacuum quark condensate becomes accessible via an analysis of the momentum dependence of the chiral limit dressed-quark mass-function.  However, for any $m_0<M_\chi$, $M_{\rm l.e.}(s,m_0)$ obtained via a linear extrapolation based on $M(s,m_1)$, $M(s,m_2)$, with $M_\chi\lsim m_1 < m_2$, must satisfy
$M_{\rm l.e.}(s,m_0)>M(s,m_0)$, where $M(s,m_0)$ is the true value of the mass-function; i.e., such an extrapolation to smaller masses must overestimate the mass-function.  That overestimate will be largest on the transition domain and hence the extrapolation will necessarily lead to an inferred value of the vacuum quark condensate that is too large.  

In any study of hadron bound states it is desirable to preserve the Ward-Takahashi identities, such as those, e.g., that are associated with the chiral and electromagnetic currents.  The usual axial-vector Ward-Takahashi identity is an expression of chiral symmetry and its breaking in QCD.  It relates the three-point axial-vector vertex to the dressed-quark propagator and the three-point pseudoscalar vertex.  The three-point functions are obtained as solutions of inhomogeneous Bethe-Salpeter equations.  The identity is therefore a statement that in QCD the kernel of the Bethe-Salpeter equations is intimately related to that of the gap equation.  In order that the features of chiral symmetry and its breaking are truly expressed in any study, that relationship must be maintained without \textit{fine tuning}.  

A truncation is necessary in order to arrive at a tractable DSE study.  At least one nonperturbative, symmetry preserving truncation scheme exists \cite{axeltrunc,hermantrunc}.  Its practical implementation requires, e.g.,  that the diagrammatic content of the dressed-quark-gluon vertex be explicitly enumerable.  This is a feature of Eq.\,(\ref{G2C}).  Hence an internally consistent and symmetry preserving study of bound states is also possible using the model defined by Eqs.\,(\ref{G2C})--(\ref{mnmodel}).  

\begin{table}[tb] 
\caption{\label{currentquark} Current-quark masses required to reproduce the experimental masses of the vector mesons.   All masses are listed in GeV.  (Adapted from Ref.\,\protect\cite{mandarvertex}.)\vspace*{0.5ex}
}
%
%
\begin{tabular*} 
{\hsize} {l@{\extracolsep{0ptplus1fil}} 
|l@{\extracolsep{0ptplus1fil}}
|l@{\extracolsep{0ptplus1fil}}} 
%
 & $0^-$(mass) & $1^-$(mass) \\\hline
$m_{u,d}=$ 0.01 & $\pi$(0.14) & $\rho$(0.77) \\
$m_s=$ 0.166 & $0^-_{s\bar s}$(0.63) & $\phi$(1.02) \\
$m_c =$ 1.33  & $\eta_c$(2.97) & $J/\psi$(3.10)\\
$m_b=$ 4.62 & $\eta_b$(9.42) & $\Upsilon_{1S}$(9.46)\\\hline
\end{tabular*} 
\end{table} 

\begin{figure}[t] 
 \centerline{\includegraphics[width=0.43\textwidth]{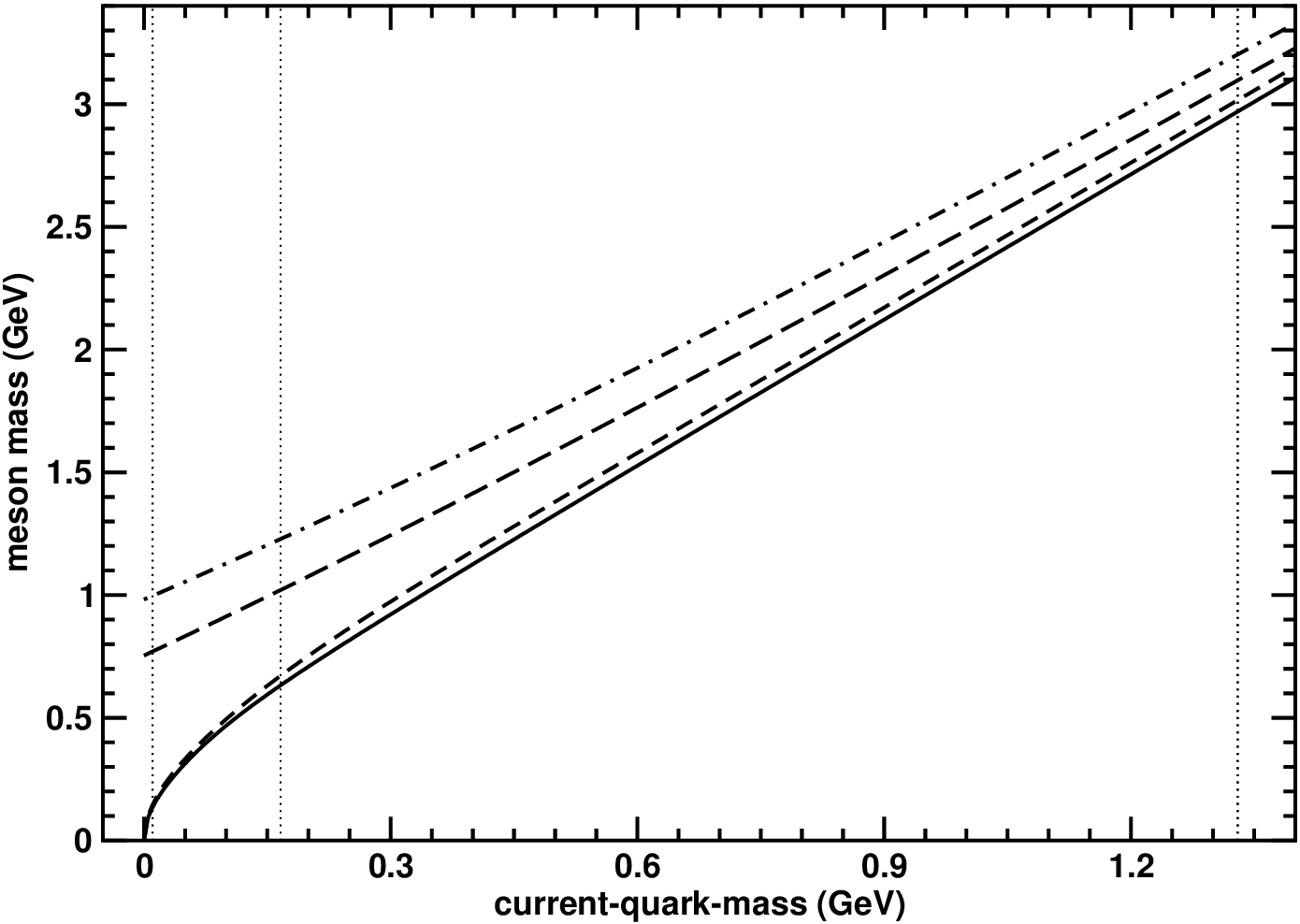}}
 
\caption{\label{massplot} Evolution of meson masses with the current-quark mass.  Solid line: pseudoscalar meson trajectory obtained with the full vertex of Eq.\,(\ref{G2C}) in the gap equation and the vertex-consistent Bethe-Salpeter kernel; short-dash line: this trajectory calculated in rainbow-ladder truncation.  Long-dash line: vector meson trajectory obtained with the full vertex and consistent Bethe-Salpeter kernel; dash-dot line: rainbow-ladder truncation result for this trajectory.  The dotted vertical lines mark the current-quark masses in Table \protect\ref{currentquark}.  (Adapted from Ref.\,\protect\cite{mandarvertex}.)}
\end{figure} 

\begin{figure}[t] 
\centerline{\includegraphics[width=0.44\textwidth]{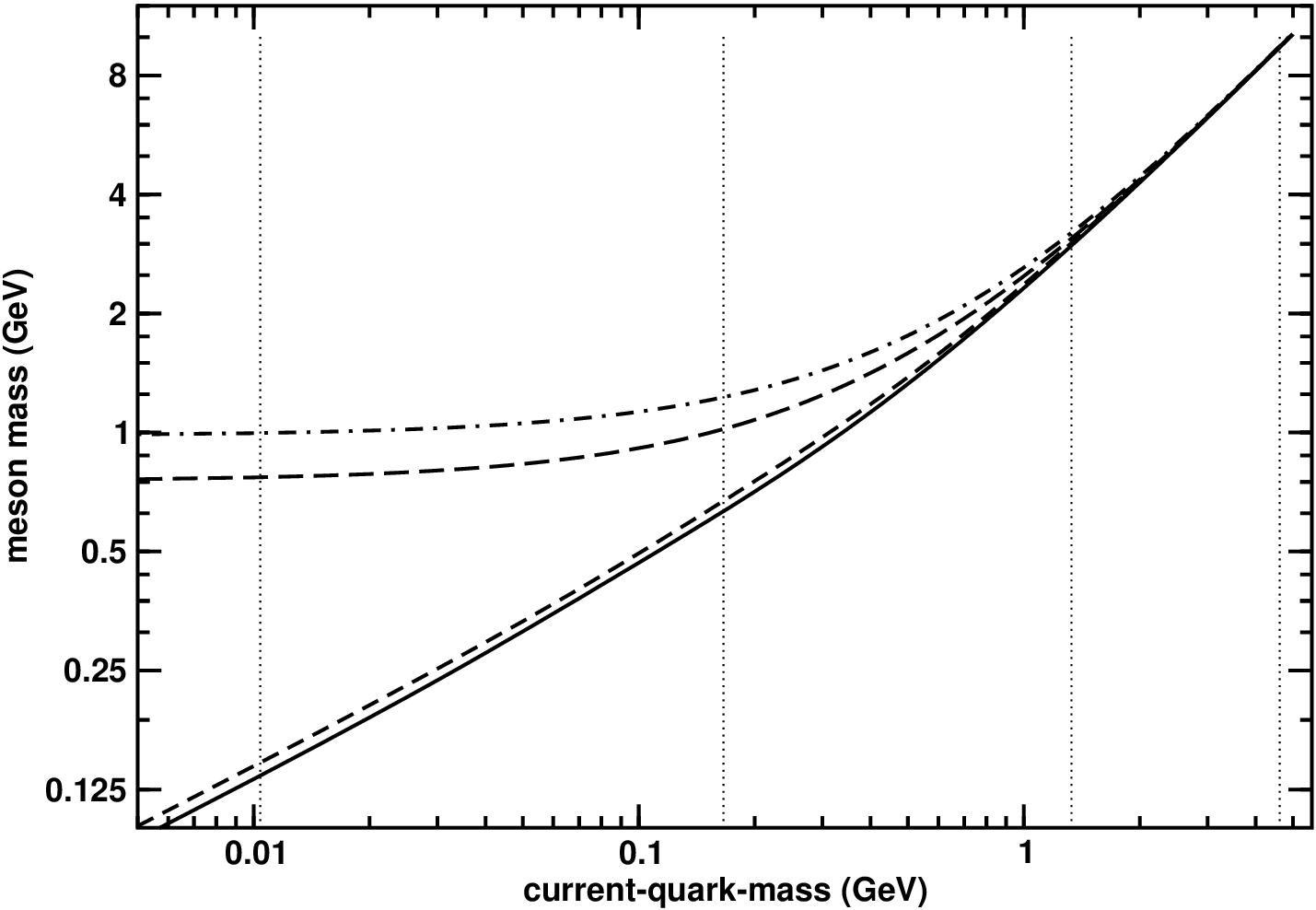}}
 
\caption{\label{massplot2} An alternative representation of the results in Fig.\,\ref{massplot}, which exposes the behaviour at large current-quark mass.  (Adapted from Ref.\,\protect\cite{mandarvertex}.)}
\end{figure} 

An extensive analysis of the ground states in the vector and flavour-nondiagonal pseudoscalar channels has been presented \cite{mandarvertex}.   The current-quark masses required to fit known vector meson masses are given in Table~\ref{currentquark}, along with the mass of the associated pseudoscalar mesons.\footnote{NB.\ $0^-_{s\bar s}$ is a fictitious pseudoscalar meson composed of unlike-flavour quarks with mass $m_s$, which is included for comparison with other nonperturbative studies.}  The values of $m_{\eta_c}$, $m_{\eta_b}$ are predictions.  Experimentally \protect\cite{pdg}, $m_{\eta_c}=2.9797\pm 0.00015\,$ and $m_{\eta_b}=9.30\pm 0.03$.  Although this model is ultraviolet finite and hence the current-quark masses cannot be directly compared with any current-quark mass-scale in QCD, the values in Table~\ref{currentquark} are quantitatively consistent with the pattern of flavour-dependence in the explicit chiral symmetry breaking mass-scales of QCD. 

The current-quark-mass-dependence of the pseudoscalar and vector meson masses was also studied in Ref.\,\cite{mandarvertex}.  This is of use in comparison with other methods and models.  The results are depicted in Figs.\,\ref{massplot}, \ref{massplot2}.  In the neighbourhood of the chiral limit the vector meson mass is approximately independent of the current-quark mass whereas the pseudoscalar meson mass increases rapidly, according to (in GeV)
\begin{equation}
\label{gmor}
m_{0^-}^2 \approx 1.33\, m\,,\,\,\; m\ll {\cal G}\,,
\end{equation}
thereby reproducing the pattern of QCD \cite{mrt98}.  

The curvature in the pseudoscalar trajectory persists over a significant domain of current-quark mass.  For example, consider two pseudoscalar mesons, one composed of unlike-flavour quarks each with mass $2 m_s$ and another composed of such quarks with mass $m_s$.  In this case
\begin{equation}
\frac{m^2_{0^-_{2 m_s}}}{m^2_{0^-_{m_s}}} = 2.4\,,
\end{equation}
a result which indicates that the nonlinear evolution exhibited in Eq.\,(\ref{gmor}) is still dominant for current-quark masses as large as twice that of the $s$-quark.  

The mode of behaviour just described is overwhelmed when the current-quark mass becomes large: $m\gg {\cal G}$.  In this limit the vector and pseudoscalar mesons become degenerate, with the mass of the ground state pseudoscalar meson rising monotonically to meet that of the vector meson.  In this model 
\begin{equation}
\left.\frac{m_{1^-}}{m_{0^-}}\right|_{m=m_c} = 1.04\,,
\end{equation}
with a splitting of $130\,$MeV, and the splitting drops to just $40\,$MeV at $m_b$; viz., only 5\% of its value in the chiral limit.  In addition to the calculated value, the general pattern of these results argues for the mass of the pseudoscalar partner of the $\Upsilon(1S)$ to lie above $9.4\,$GeV.  Indeed, one expects the mass splitting to be much less than $m_{J/\psi}-m_{\eta_c}$, not more.

\begin{figure}[t] 
\centerline{\includegraphics[width=0.45\textwidth]{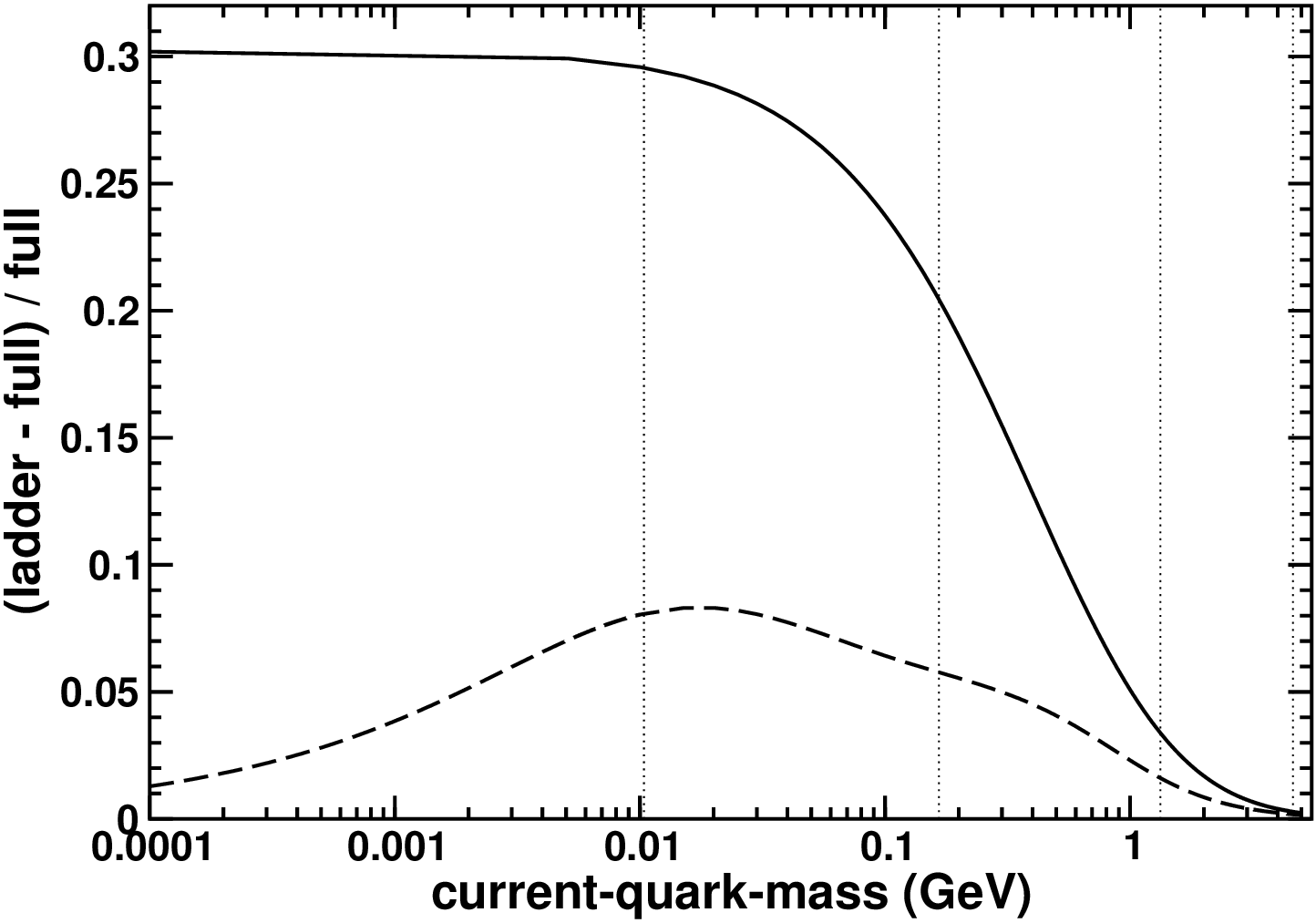}}

\caption{\label{ladderfull} Evolution with current-quark mass of the relative difference between the meson mass calculated in the rainbow-ladder truncation and the exact value; namely, that obtained using the full vertex of Eq.\,(\ref{G2C}) in the gap equation and the vertex-consistent Bethe-Salpeter kernel.  Solid lines: vector meson trajectories; and dashed-lines; pseudoscalar meson trajectories.  The dotted vertical lines mark the current-quark masses in Table \protect\ref{currentquark}.  (Adapted from Ref.\,\protect\cite{mandarvertex}.)}
\end{figure} 

Figure~\ref{ladderfull} shows that with growing current-quark mass the rainbow-ladder truncation provides an increasingly accurate estimate of the ground state vector meson mass.  At the $s$-quark mass the relative error is 20\% but that has fallen to $< 4$\% at the $c$-quark mass.  

Similar statements are true in the valid pseudoscalar channels.  In fact, in this case the agreement between the truncated and exact results is always better; e.g., the absolute difference reaches its peak of $\approx 60\,$MeV at $m\sim 4\,m_s$ whereat the relative error is only 3\%.  This behaviour is fundamentally because of Goldstone's theorem, which requires that all legitimate truncations preserve the axial-vector Ward-Takahashi identity and hence give a massless pseudoscalar meson in the chiral limit.  It is practically useful, too, because it indicates that the parameters of a model meant to be employed in a rainbow-ladder truncation study of hadron observables may reliably be fixed by fitting to the values of quantities calculated in the neighbourhood of the chiral limit.

The general conclusion suggested by Fig.\,\ref{ladderfull} is that with increasing current-quark mass the contributions from nonplanar diagrams and vertex corrections are suppressed in both the gap and Bethe-Salpeter equations.\footnote{Naturally, they must still be included in precision spectroscopic calculations at real current-quark masses.}  The implication for QCD is that the RGI rainbow-ladder truncation is exact in the static source limit.  

This is the anticipated outcome, with which studies such as Ref.\,\cite{davydichev} are consistent.  There are no extant counterexamples.  If this truly is the case then a linear potential at long-range between static sources should be expressed in the running coupling constant that defines the RGI rainbow-ladder truncation; \textit{\`{a} la} Richardson's potential \cite{richardson}, for example.  According to the interpretation of Ref.\,\cite{mm97}, this is a property of the RGI rainbow-ladder model introduced in Refs.\,\cite{mrtmodel} and used with success in the study of hadron observables \cite{pieterrev}.  

Elucidating the nature of the interaction between quarks at long-range is a contemporary challenge in hadron physics.  Information is contained in the hadron spectrum; e.g., in the properties of the radial excitations of ground state mesons \cite{andreas}, axial vector mesons \cite{a1b1}, and exotics and missing nucleon resonances \cite{pdg}.  The methods sketched herein are well suited to extracting that information. 

\medskip
\hspace*{-\parindent}{\bf Acknowledgments}.\ We acknowledge valuable communications with P.\,C.\ Tandy.  CDR thanks the organisers and the staff of the CSSM for their hospitality and support.  
This work was also supported by: %
the Austrian Research Foundation \textit{FWF, 
Erwin-Schr\"odinger-Stipendium} no.\ J2233-N08; Department of Energy, 
Office of Nuclear Physics, contract no.\ W-31-109-ENG-38; National Science Foundation contract no.\ INT-0129236; the \textit{A.\,v.\ 
Humboldt-Stiftung} via a \textit{F.\,W.\ Bessel Forschungspreis};  and benefited from the facilities ANL's Computing Resource Center.

\end{document}